\documentclass{pasj00}
\begin{document}
\SetRunningHead{N. Isobe, H. Seta \& M.S. Tashiro}
{Electron and magnetic energy densities in the east lobe of DA 240}
\Received{2011/03/23}
\Accepted{2011/05/17}

\title{Suzaku measurement of electron and magnetic energy densities 
in the east lobe of the giant radio galaxy DA 240}

\author{%
Naoki       \textsc{Isobe}      \altaffilmark{1,2},
Hiromi      \textsc{Seta}       \altaffilmark{3},
\&
Makoto  S.  \textsc{Tashiro}    \altaffilmark{3}
}
\altaffiltext{1}{Department of Astronomy, Kyoto University, \\
        Kitashirakawa-Oiwake-cho, Sakyo-ku, Kyoto 606-8502, Japan}
\altaffiltext{2}{
        Institute of Space and Astronautical Science (ISAS), 
        Japan Aerospace Exploration Agency (JAXA) \\ 
        3-1-1 Yoshinodai, Chuo-ku, Sagamihara, Kanagawa 252-5210, Japan}
\email{n-isobe@ir.isas.jaxa.jp}
\altaffiltext{3}{Department of Physics, Saitama University, \\
        255 Shimo-Okubo, Sakura-ku, Saitama, 338-8570, Japan}
\maketitle
\begin{abstract}
A careful analysis of the Suzaku data of the giant radio galaxy DA 240,
of which the size is $1.48$ Mpc,  
revealed diffuse X-ray emission associated with its east lobe.
The diffuse X-ray spectrum was described with a simple power-law model 
with a photon index of $\Gamma = 1.92_{-0.17}^{+0.13}$$_{-0.06}^{+0.04}$,
where the first and second errors represent 
the statistical and systematic ones. 
The agreement with the synchrotron radio photon index, 
$\Gamma_{\rm R} = 1.95 \pm 0.01 $ in 326 -- 608.5 MHz,
ensures that the excess X-ray emission is attributed to 
the inverse Compton emission from the synchrotron-radiating electrons,
boosting up the cosmic microwave background photons. 
From the X-ray flux density,
$51.5\pm3.9_{-5.4}^{+6.2}$ nJy at 1 keV 
derived with the photon index fixed at $\Gamma_{\rm R}$, 
in comparison with the synchrotron radio intensity of 
$10.30 \pm 0.12$ Jy at 326 MHz,
the magnetic and electron energy densities was estimated as 
$u_{\rm m} = (3.0\pm0.2\pm0.4)\times 10^{-14}$ ergs cm$^{-3}$
and
$u_{\rm e} = (3.4_{-0.2}^{+0.3}$$_{-0.4}^{+0.5})\times10^{-14}$ ergs cm$^{-3}$
integrated over the electron Lorentz factor of $10^3$ -- $10^5$, respectively.
Thus, the east lobe is found to reside in an equipartition condition
between the electrons and magnetic field parametrized 
as $u_{\rm e}/u_{\rm m} = 1.1_{-0.1}^{+0.2}$$_{-0.2}^{+0.4}$. 
The east lobe of DA 240 is indicated 
to exhibit the lowest value of $u_{\rm e}$,
among all the X-ray detected lobes of radio galaxies. 
A comparison of the energetics in the giant radio galaxies 
with a size of $\sim 1$ Mpc to those in the smaller objects 
suggests a possibility that radio galaxies lose their jet power, 
as they evolve from $\sim100$ kpc to $\sim1$ Mpc.
\end{abstract}

\section{Introduction} 
\label{sec:intro}
Among extragalactic radio source,
those with a physical size larger than $\sim 1$ Mpc 
at the source rest frame are called as {\it giant radio galaxies}. 
The synchrotron aging technique suggests that 
they are relatively old sources 
with a typical age of $\sim 100$ Myr \citep{giants_age}.  
Therefore, from the giant radio galaxies,
we can explore 
the late phase of the evolution of the radio sources 
and the associated jets, which is not yet well understood. 

The bulk kinetic energy, carried by the jets of a radio galaxy, 
is integrated within the lobes 
after it is randomized and converted into the particle and magnetic energies 
at its terminal hot spots. 
The lobes are regarded as an important indicator of 
the past activity of the jets.
One of the most valuable tools to measure the electron and magnetic energies 
stored in the lobes is diffuse inverse Compton (IC) X-ray emission, 
where the cosmic microwave background (CMB) photons are up-scattered
(\cite{CMB_IC}; hereafter IC/CMB). 
Actually, 
the energetics in the lobes of numbers of radio galaxies are investigated from 
the diffuse X-ray emission 
which was naturally interpreted as the IC/CMB emission
in the last two decades, with ROSAT \citep{ForA_ROSAT}, 
ASCA (e.g., \cite{ForA_ASCA,CenB,ForA_ASCA_2}),
Chandra (e.g., \cite{3C452,lobes_Croston,CygA_Yaji}) 
and XMM-Newton (e.g., \cite{3C98,ForA}). 
These results indicate a dominance of the electron energy 
over the magnetic one by typically an order of magnitude
in the lobes of radio galaxies (\cite{3C35} for a summary).
However, most of the sample is limited to the sources
with a total dimension of $D\lesssim 500$ kpc, so far. 

\begin{figure*}[t]
\begin{center}
\FigureFile(80mm,80mm){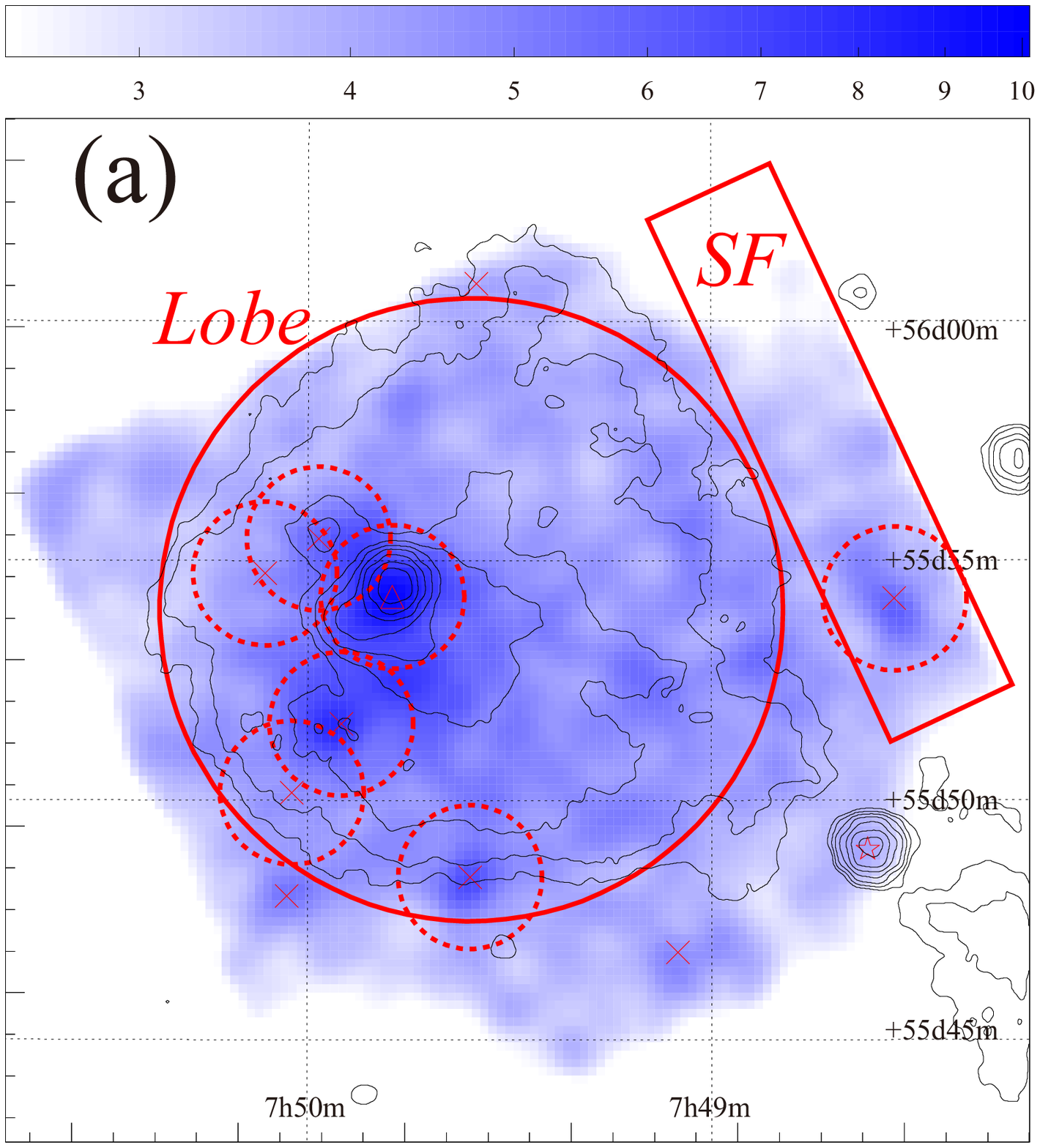}
\hspace{5mm}
\FigureFile(78.5mm,78.5mm){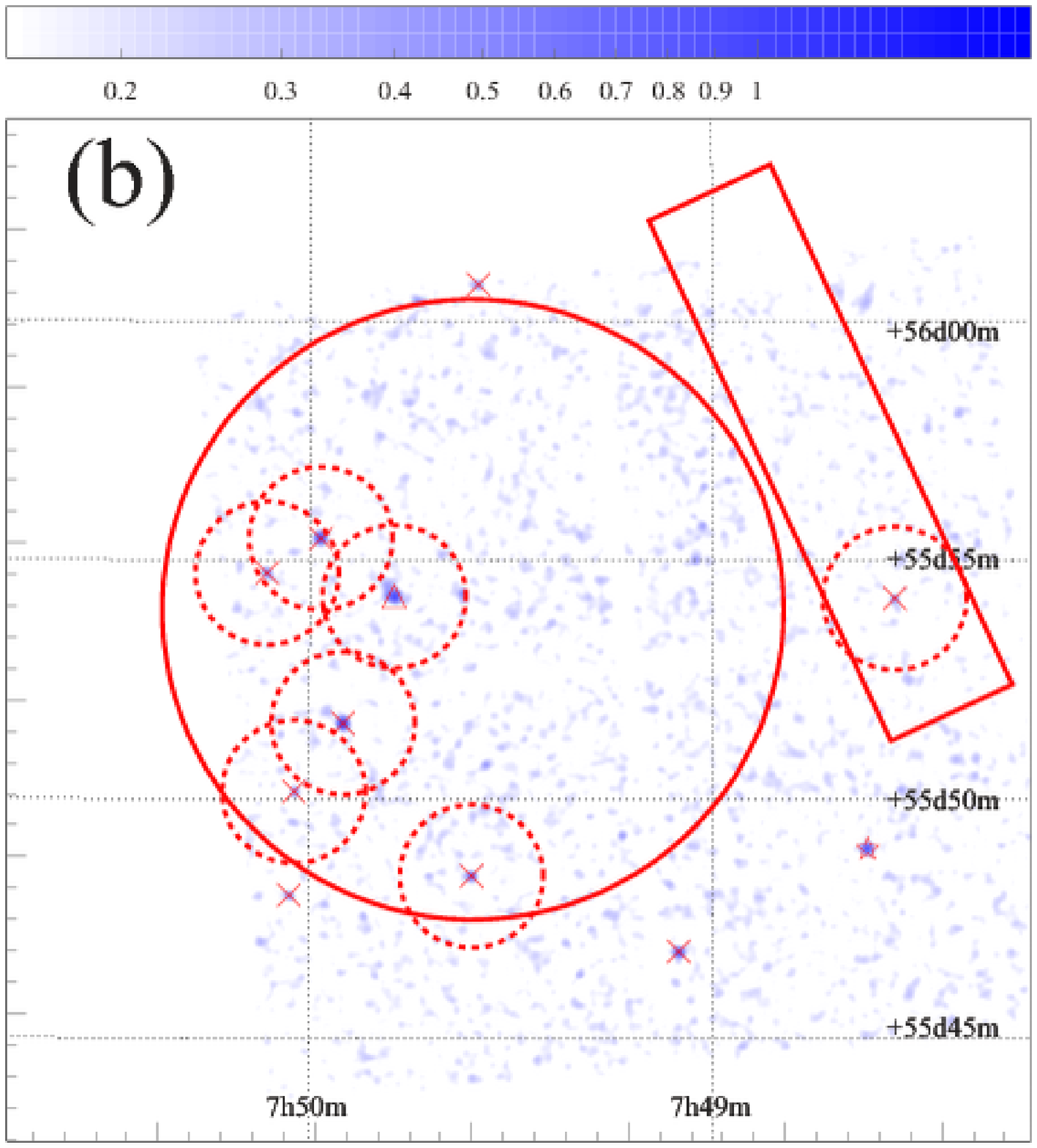}
\end{center}
\caption{(a) Suzaku XIS image of the east lobe of DA 240 
in the 0.5 -- 10 keV range,
heavily smoothed with a two-dimensional Gaussian function of a $20''$ radius.
The background events were not subtracted, 
while the exposure was not corrected. 
The scale bar on the top indicates 
the X-ray counts integrated within a $10'' \times 10''$ bin.  
A 608.5 MHz radio image (Leahy et al. unpublished)
is superposed with contours.
The Chandra position of the nucleus and hot spot 
are shown with the star and triangle, respectively. 
The other Chandra X-ray point sources, 
listed in table \ref{table:CXO_Src}, are plotted with the crosses.
The XIS events of the east lobe and XRB were 
accumulated within the solid circle denoted as {\bf Lobe}
and solid rectangle labeled as {\bf SF}, respectively, 
with those in the dashed circles removed. 
(b) Chandra ACIS image of the same sky field in 0.3 -- 10 keV,
smoothed with a two-dimensional Gaussian function of a $4''$ radius.
The scale bar displays 
the X-ray counts integrated within a $2'' \times 2''$ bin. 
The position of the Chandra X-ray sources 
and the spectral integration regions for the XIS 
are shown in the manner similar to that in panel (a).}
\label{fig:image}
\end{figure*}

The X-ray Imaging Spectrometer (XIS; \cite{XIS}) 
onboard the Suzaku observatory \citep{Suzaku},
in combination with the Hard X-ray Detector (HXD; \cite{HXD}), 
opened a new window for the X-ray study of the lobes of radio galaxies
(e.g., \cite{ForA_Suzaku}),
thanks to its low and stable instrumental background,
which is crucial to investigate diffuse X-ray emission. 
With the Suzaku XIS,
detections of the diffuse X-ray photons has subsequently been reported 
from the giant radio galaxies 3C 326 \citep{3C326} and 3C 35 \citep{3C35},
and these X-rays were successfully attributed to be radiated 
via the IC/CMB process. 
Based on these results, \citet{3C35} proposed a possibility that 
the current jet power of the giant radio galaxies are significantly lower 
than that of the smaller objects. 
This motivated us to make a systematic X-ray study of giant radio galaxies. 
 
The giant radio galaxy DA 240
is located at a redshift of $z = 0.035661 $ \citep{redshift}.
It is optically classified into 
a low-excitation radio galaxy \citep{RG_summary}
with an elliptical host \citep{DA240_host}.  
Its radio images (e.g., \cite{GRG}) revealed 
its classical Fanaroff-Riley (FR) II morphology,
where the lobes have a relatively flat brightness distribution.
The total angular size of the source, \timeform{35.2'}, 
corresponds to a physical size of $1.48$ Mpc at the redshift of the source. 
Its spectral age was estimated as $56$ Myr \citep{DA240_age}, 
suggesting that it is a relatively old source. 
The object has a high integrated synchrotron radio flux 
of $9.2$ Jy at 609 MHz \citep{GRG},
which suggests a high IC/CMB X-ray flux from its lobes. 
Although there is a bright radio hot spot in the east lobe,
which was also detected in X-rays with XMM-Newton \citep{DA240_HS},
its contamination to the lobe is to be estimated 
and to be clearly removed from the diffuse lobe emission 
owing to the large size of the lobe 
in comparison with the point spread function 
of the X-ray telescope (XRT; \cite{XRT}) over the XIS  
with a half power diameter of $\sim$\timeform{2'} \citep{XRT}.
These make DA 240 suitable for a Suzaku observation. 

In the present paper, 
we adopted the cosmology with 
$H_{\rm 0} = 71$ km s$^{-1}$ Mpc$^{-1}$, 
$\Omega_{\rm m} = 0.27$, and  $\Omega_{\lambda} = 0.73$.
These give the luminosity distance of $ 154.8 $ Mpc
and the angle-to-size conversion ratio 
of $41.97$ kpc/$1'$, at the redshift of DA 240 ($z = 0.035661$).

\section{Observation and Data Reduction} 
\subsection{Suzaku Observation} 
\label{sec:suzaku_obs}
We performed a Suzaku observation of 
the giant radio galaxy DA 240, on 2010 March 19 -- 21.
We pointed the telescope at the east lobe,
since its integrated radio flux is higher than that of the west lobe.
In order to minimize the influence of the anomalous columns of XIS 0 
\footnote{{\tt http://www.astro.isas.jaxa.jp/suzaku/doc/suzakumemo/suzakumemo-2010-01.pdf}}
and the hole of the optical blocking filter over XIS 1
\footnote{{\tt http://www.astro.isas.jaxa.jp/suzaku/doc/suzakumemo/suzakumemo-2010-03.pdf}}, 
the sky position with the J2000 coordinate of  
($\alpha$, $\delta$) = (\timeform{117.3696\circ}, \timeform{+55.8964\circ}) 
was placed at the XIS nominal position (\cite{XRT}).
As a result, the whole east lobe was observed 
within the clean field of view of the XIS.  
The normal clocking mode with no window option was adopted 
for the XIS 
while the normal mode was employed for the HXD.

We reduced the data with the latest version of 
the standard software package HEADAS 6.10. 
Since the lobe emission was found to be too faint for the HXD, 
only the XIS data are utilized.
All the XIS data were reprocessed, 
by referring to the CALDB updated at 2010 July 30. 
We screened the data under the following standard criteria; 
the spacecraft is outside the south Atlantic anomaly (SAA),
the time after an exit from the SAA is larger than 436 s,
the geometric cut-off rigidity is higher than $6$ GV, 
the source elevation above the rim of bright and night Earth is 
higher than $20^\circ$ and $5^\circ$, respectively, 
and the XIS data are free from telemetry saturation. 
These yielded a good exposure of $75.5$ ks.
In the scientific analysis below,
we selected the XIS events with a grade of 0, 2, 3, 4, or 6.

\subsection{Archival Chandra Data} 
In order to identify possible contaminating faint X-ray sources
unresolved with the XIS angular resolution, 
we analyzed the archival Chandra data. 
The sky field similar to that in the Suzaku observation 
was observed with the Chandra ACIS-I on 2009 January 12 
(ObsID $ = 10237$).
The latest software package CIAO 4.3 was utilized 
in combination with CALDB 4.4.1. 
We reprocessed the ACIS data to create the new level 2 event file
with the tool {\tt chandra\_repro},
following the standard manner. 
Because no significant background variation was found 
throughout the observation,
no additional screening was performed to the new level 2 event file. 
As a result, a good exposure of $24.1$ ks was derived. 
We adopted the grade selection same as for the XIS (i.e., 0, 2, 3, 4, or 6). 

\begin{table}[t]
\caption{Chandra X-ray sources in the XIS field of view.}
\label{table:CXO_Src}
\begin{center}
\begin{tabular}{llll}
\hline\hline 
($\alpha$, $\delta$)                       & $\Delta \theta$ \footnotemark[$*$] 
                                                 & $\sigma$ \footnotemark[$\dagger$] & Source ID \\  
\hline 
(\timeform{117.2709\circ}, \timeform{+55.7804\circ}) & 0.6 & 11.3 & \\ 
(\timeform{117.5127\circ}, \timeform{+55.8000\circ}) & 1.0 &  5.7 & \\ 
(\timeform{117.3994\circ}, \timeform{+55.8067\circ})\footnotemark[$\ddagger$]
                                           & 0.5 &  8.8 & \\ 
(\timeform{117.1535\circ}, \timeform{+55.8162\circ}) & 0.6 & 11.9 & The nucleus \\ 
(\timeform{117.5097\circ}, \timeform{+55.8360\circ})\footnotemark[$\ddagger$]
                                           & 1.1 &  5.7 & \\ 
(\timeform{117.4792\circ}, \timeform{+55.8600\circ})\footnotemark[$\ddagger$]
                                           & 0.5 & 17.3 & \\ 
(\timeform{117.1364\circ}, \timeform{+55.9035\circ})\footnotemark[$\S$]
                                           & 0.5 &  5.1 & \\ 
(\timeform{117.4477\circ}, \timeform{+55.9041\circ})\footnotemark[$\ddagger$]
                                           & 0.4 & 13.0 & The hot spot \\ 
(\timeform{117.5268\circ}, \timeform{+55.9122\circ})\footnotemark[$\ddagger$]
                                           & 0.9 &  6.7 & \\ 
(\timeform{117.4933\circ}, \timeform{+55.9244\circ})\footnotemark[$\ddagger$]
                                           & 0.4 & 14.0 & Ark 141 \\ 
(\timeform{117.3957\circ}, \timeform{+56.0130\circ}) & 1.0 &  5.8 & \\ 
\hline 
\multicolumn{4}{@{}l@{}}{\hbox to 0pt{\parbox{80mm}{\footnotesize
\par\noindent
\footnotemark[$*$] Position error in arcsec. 
\par\noindent
\footnotemark[$\dagger$] Source significance. 
\par\noindent
\footnotemark[$\ddagger$] Removed from the lobe spectrum.  
\par\noindent
\footnotemark[$\S$] Removed from the SF spectrum.  
}\hss}}
\end{tabular}
\end{center}
\end{table}

\section{Results} 
\label{sec:results}
\subsection{X-ray image}
\label{sec:image}
The panel (a) of figure \ref{fig:image} displays the background-inclusive 
$0.5$ -- $10$ keV Suzaku XIS image of the east lobe of DA 240, 
on which a 608.5 MHz radio image 
(Leahy et al., unpublished)\footnote{
Taken from ``An Atlas of DRAGNs'', edited by Leahy, Bridle, \& Strom;
{\tt http://www.jb.man.ac.uk/atlas/}.}  
is over-plotted with contours. 
We smoothed the XIS image with 
a two-dimensional Gaussian kernel of a $20''$ radius.
The regions irradiated by the radioactive calibration source ($^{55}$Fe) 
on the corners of the XIS chips were removed, 
while the exposure correction was not performed. 
We noticed a systematic offset of $16.3''$ 
in the XIS coordinate determination,
while this offset is found to be 
within the current systematic uncertainties \citep{Suzaku_pos_acc}. 
We thus calibrated the XIS coordinate by ourselves,
by referring to the Chandra position (see below)
of the hot spot in the east lobe (the triangle in figure \ref{fig:image}), 
which is clearly detected in the XIS image.
After the coordinate correction,  
figure \ref{fig:image} indicates that 
the X-ray position of the hot spot deviates 
from its radio position by $\sim 10''$ toward the south direction. 
The similar positional offset,  
in addition to the detailed spatial structure of the hot spot,
is reported by \citet{DA240_HS},
by using the higher-resolution radio and X-ray images 
obtained with the Very Large Array and XMM-Newton, respectively. 
We regard the detailed examination on this point 
is beyond the scope of the present paper.

Most of the east lobe was confirmed to be contained 
within the clean XIS field of view,
while the nucleus of DA 240,
indicated by the star in figure \ref{fig:image},  
was found to fall unfortunately on the XIS 0 anomalous columns.  
From the image alone, it is unclear whether or not diffuse faint 
X-ray emission associated with the lobe was detected. 

The XIS image suggests several X-ray sources within the east lobe.
Then, the Chandra ACIS image of the similar sky field 
in the $0.3$ -- $10$ keV range is shown in figure \ref{fig:image} (b). 
The ACIS field of view is found to cover a large part of the east lobe.
We made use of the CIAO tool {\tt wavdetect} 
in order to pick up contaminating faint X-ray sources. 
Table \ref{table:CXO_Src} lists the detected X-ray sources 
with a significance higher than $5 \sigma$,
among which the nucleus and hot spot of DA 240 were included. 
These sources are indicated on figure \ref{fig:image} with crosses. 

\begin{figure}[t]
\begin{center}
\includegraphics[angle=-90,width=8cm]{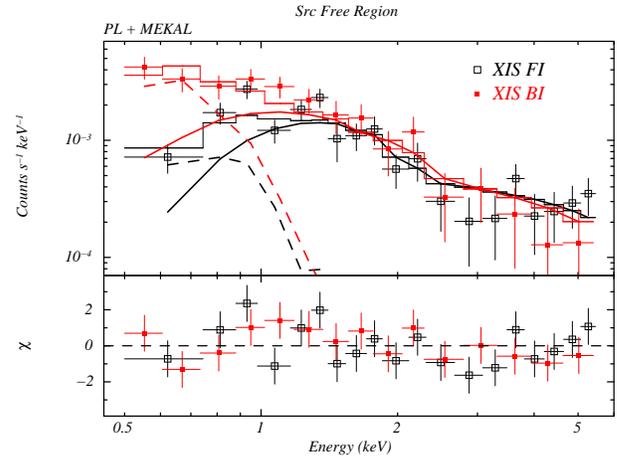}
\end{center}
\caption{Suzaku XIS spectrum of the SF region, 
fitted with the PL+MEKAL model. The solid and dashed lines indicate 
the PL and MEKAL components, respectively.}
\label{fig:SrcFree}
\end{figure}

\begin{figure*}[t]
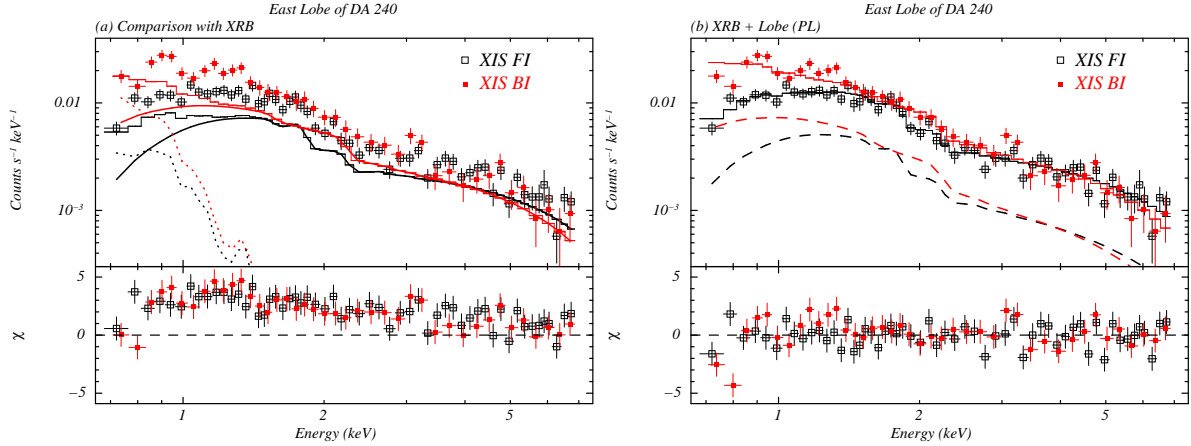

\begin{center}
\includegraphics[angle=-90,width=7.8cm]{fig3a.ps}
\includegraphics[angle=-90,width=7.8cm]{fig3b.ps}
\end{center}
\caption{Suzaku XIS spectrum of the east lobe of DA 240, 
accumulated from the Lobe region shown in figure \ref{fig:image}.
Panel (a) compares the data with the XRB model spectrum 
in the field of DA 240, determined from the SF region 
(figure \ref{fig:SrcFree} and table \ref{table:SrcFree}).
The solid and dashed lines indicate the PL and MEKAL components, respectively. 
In panel (b), the best-fit model, 
in which the additional PL component (the dashed lines)
describing the X-ray emission from the east lobe was taken into account. 
For clarity, the XRB model is omitted. }
\label{fig:lobe}
\end{figure*}

\subsection{X-ray background in the field}
For the spectral analysis of diffuse and faint X-ray sources, 
it is of crucial importance to determine 
accurately the level of the X-ray background (XRB). 
Hence, we first evaluated the XRB spectrum in the field,
from the source free (SF) region within the XIS field of view, 
which is displayed with the solid rectangle 
denoted as {\bf SF} in figure \ref{fig:image}. 
The region has a size of \timeform{12'} $\times$ \timeform{2'.8}.
Because figure \ref{fig:image} (b) reveals 
a faint X-ray source detected with the Chandra ACIS 
within the SF region (see table \ref{table:CXO_Src}), 
we rejected 
a circle with a radius of \timeform{1.5'} 
centered on the source (the dashed circle in figure \ref{fig:image}).
We confirmed that a source rejection with a larger radius 
yielded a similar result within the statistical uncertainties.

The non-X-ray background (NXB) in the region was estimated 
by the HEADAS tool {\tt xisnxbgen}. 
The tool is reported to reproduce the NXB spectrum 
in the $1$ -- $7$ keV range, with an accuracy better than $\sim 3$\%
for a typical exposure of $50$ ks \citep{xisnxbgen}.
Figure \ref{fig:SrcFree} shows the NXB-subtracted XIS spectrum 
of the SF region. 
Although the region irradiated 
by the radioactive calibration source ($^{55}$Fe)
in the north part of the SF region 
for the front-illuminated (FI) CCD chips 
(XIS 0 and 3; \cite{XIS}) was not removed,  
we limited the FI data below $5.5$ keV. 

Referring to \citet{XRB_ASCA} and \citet{XRB_XMM}, 
we approximate the XIS spectrum of the SF region 
with a composite model consisting of a hard power-law (PL) component
and a soft thermal plasma emission described by the MEKAL \citep{MEKAL} code.  
The PL component is thought to be composed of unresolved faint X-ray sources,
including distant active galactic nuclei, 
while 
the soft thermal emission is considered to be of Galactic origin. 
The PL photon index was fixed at $\Gamma = 1.41$ \citep{XRB_ASCA},
while the temperature of the thermal MEKAL component was left free 
and the solar abundance ratio was adopted. 
We assumed a common photoelectric absorption 
for both components 
with the Galactic hydrogen column density toward DA 240
($N_{\rm H} = 4.9 \times 10^{20}$ cm$^{-1}$; \cite{NH}). 
Response matrix functions (rmf) 
of the XIS was generated by the tool {\tt xisrmfgen}.
Assuming a uniform diffuse X-ray source with a $20'$ radius, 
we calculated the auxiliary response files (arf)
by the tool {\tt xissimarfgen} \citep{xissimarf}. 

We obtained an acceptable fit ($\chi^2/{\rm dof} = 33.2 / 31$)
by the PL+MEKAL model, as shown with the histograms 
in figure \ref{fig:SrcFree}.
The best-fit spectral parameters are 
summarized in table \ref{table:SrcFree}. 
The total absorption-inclusive surface brightness of the XRB was 
measured as 
$f = 7.0_{-0.6}^{+0.5} \times 10^{-8}$ ergs s$^{-1}$ cm$^{-2}$ str$^{-1}$
in the 0.5 -- 10 keV range.
The MEKAL temperature derived as $kT = 0.20_{-0.03}^{+0.06}$ keV
is reasonable for the diffuse Galactic background emission
($kT = 0.204 \pm 0.009$ keV; \cite{XRB_XMM}).  
The $2$ -- $10$ keV surface brightness of the PL component, 
$f_{\rm PL} = (4.4 \pm 0.5) \times 10^{-8}$ ergs s$^{-1}$ cm$^{-2}$ str$^{-1}$,
was found to be consistent with the result in \citet{XRB_ASCA}
within the $3\sigma$ uncertainty. 
Therefore, we decided to adopt the best-fit PL+MEKAL model 
for the XRB spectrum to the east lobe of DA 240. 

\begin{table}[t]
\caption{Summary of the PL+MEKAL fitting to the XIS spectra of the SF region.}
\label{table:SrcFree}
\begin{center}
\begin{tabular}{ll}
\hline\hline 
Parameters                             & Values \\
\hline 
$N_{\rm H}$ (cm$^{-2}$)                & $4.9 \times 10^{20}$ \footnotemark[$*$]      \\
$\Gamma$                               & 1.41 \footnotemark[$\dagger$] \\  
$f_{\rm PL}$ (ergs s$^{-1}$ cm$^{-2}$ str$^{-1}$)\footnotemark[$\ddagger$] 
                                       & $(4.4 \pm 0.5) \times 10^{-8}$\\
$kT$ (keV)                             & $0.2_{-0.3}^{+0.6}$\\ 
$f_{\rm th}$ (ergs s$^{-1}$ cm$^{-2}$ str$^{-1}$)\footnotemark[$\S $]      
                                       & $(1.2 \pm 0.5) \times 10^{-8}$\\
$\chi^2/{\rm dof}$                     & $33.2/31$ \\ 
\hline 
\multicolumn{2}{@{}l@{}}{\hbox to 0pt{\parbox{80mm}{\footnotesize
\par\noindent
\footnotemark[$*$] Fixed at the Galactic value \citep{NH}.
\par\noindent
\footnotemark[$\dagger$] Taken from \citet{XRB_ASCA}.
\par\noindent
\footnotemark[$\ddagger$] Absorption-inclusive surface brightness 
                          of the PL component in 2 - 10 keV. 
\par\noindent
\footnotemark[$\S$] Absorption-inclusive surface brightness 
                    of the MEKAL component in 0.5 - 2 keV. 
}\hss}}
\end{tabular}
\end{center}
\end{table}

\begin{table*}[t]
\caption{Summary of signal statistics in the 0.7 -- 7 keV range}
\label{table:stats}
\begin{center}
\begin{tabular}{lllll}
\hline\hline 
Region & Component  & FI rate (cts s$^{-1}$)
                    & BI rate (cts s$^{-1}$) 
                    & $F_{\rm X}$ (ergs s$^{-1}$ cm$^{-2}$) \footnotemark[$*$] \\
\hline 
Lobe   & Data       & $(3.81 \pm 0.05)\times10^{-2}$       
                    & $(5.62\pm0.09)\times10^{-2}$                     
                    & \\
       & NXB \footnotemark[$\dagger$] & $(1.36 \pm 0.01 \pm 0.04)\times10^{-2}$ 
                    & $(2.35 \pm 0.02 \pm 0.07)\times10^{-2}$          
                    & \\
       & XRB        & $(1.63_{-0.10}^{+0.09})\times10^{-2}$  
                    & $(2.09_{-0.13}^{+0.10})\times10^{-2}$              
                    & $(4.9 \pm 0.5 ) \times 10^{-12}$ \footnotemark[$\ddagger$] \\
       & Signal \footnotemark[$\dagger$] & $(0.83 \pm 0.05 _{-0.10}^{+0.11}) \times 10^{-2}$    
                    & $(1.18 \pm 0.09 _{-0.13}^{+0.15}) \times 10^{-2}$  
                    & $(2.9\pm 0.2_{-0.3}^{+0.4}) \times 10^{-13}$ \footnotemark[$\S$] \\
\hline 
hot spot &          & $(1.77 \pm 0.21)\times 10^{-3}$
                    & $(1.82 \pm 0.23)\times 10^{-3}$
                    & $(5.3 \pm 0.5)\times 10^{-14}$ \footnotemark[$\|$] \\
\hline 
nucleus \footnotemark[$\#$] 
         &          & ---
                    & $ < 5.6 \times 10^{-4}$   
                    & $ < 2.7 \times 10^{-14}$ \footnotemark[$**$]       \\
\hline 
\multicolumn{5}{@{}l@{}}{\hbox to 0pt{\parbox{165mm}{\footnotesize
\par\noindent
\footnotemark[$*$] Absorption-inclusive 0.7 -- 7 keV flux 
\par\noindent
\footnotemark[$\dagger$] The first and second errors 
                         represent the statistical and systematic ones, respectively. 
\par\noindent
\footnotemark[$\ddagger$] Normalized to the sky region with a radius of $20$ arcmin, 
                          based on the best-fit PL+MEKAL spectrum (table \ref{table:SrcFree}).
\par\noindent
\footnotemark[$\S$] Evaluated from Case 1 in table \ref{table:lobe}.
\par\noindent
\footnotemark[$\|$] Evaluated from the best-fit PL spectrum (table \ref{table:hotspot}).
\par\noindent
\footnotemark[$\#$] The $3 \sigma$ BI upper limit is shown. 
\par\noindent
\footnotemark[$**$] A PL spectrum with a photon index of $\Gamma = 1.91$ \citep{DA240_HS}, 
                    modified by the Galactic absorption, is assumed. 
}\hss}}
\end{tabular}
\end{center}
\end{table*}

\subsection{X-ray spectrum of the east lobe} 
We extracted the XIS spectrum of the east lobe of DA 240,
from the solid circle denoted as {\bf Lobe} in figure \ref{fig:image} 
with a radius of \timeform{6.5'} 
(corresponding to 272.8 kpc at the redshift of DA 240).
The contamination from the Chandra X-ray sources within the Lobe region,
tabulated in table \ref{table:SrcFree}, was removed, 
by rejecting the dashed circles in figure \ref{fig:image},
all of which has a radius of {\timeform 1.5'}. 
Figure \ref{fig:lobe} shows the XIS spectrum of the Lobe region
in the range of $0.7$ -- $7$ keV, 
after the NXB, evaluated by the {\tt xisnxbgen}, was subtracted.
The signal statistics within the Lobe region are summarized 
in table \ref{table:stats}.
The 0.7 -- 7 keV data count rate per CCD chip was measured 
as  $(3.81 \pm 0.05) \times 10^{-2}$ cts s$^{-1}$ 
and $(5.62 \pm 0.09) \times 10^{-2}$ cts s$^{-1}$  
with the XIS FI and BI, respectively,
while the NXB count rate was estimated   
as $(1.36 \pm 0.01) \times 10^{-2}$ cts s$^{-1}$ 
and $(2.35 \pm 0.02) \times 10^{-2}$ cts s$^{-1}$. 
Here, the $1 \sigma$ statistical errors are accompanied with the value. 
We, thus, confirmed the NXB-subtracted X-ray signals of
$(2.46 \pm 0.05)\times 10^{-2}$ cts s$^{-1}$ and     
$(3.27 \pm 0.09)\times 10^{-2}$ cts s$^{-1}$,
yielding the statistical significance of $48.2 \sigma$ and $37.2 \sigma$,
with the XIS FI and BI, respectively.
In addition, these signals are found to well exceed 
the typical systematic NXB fluctuation of $ \sim 3 \% $ \citep{xisnxbgen},
which corresponds to the FI and BI count rates of 
$0.04 \times 10^{-2}$ cts s$^{-1}$ and $0.07 \times 10^{-2}$ cts s$^{-1}$
respectively.

In panel (a) of figure \ref{fig:lobe},
we compare the NXB-subtracted XIS spectrum of the Lobe region 
with the XRB in the field.
The best-fit PL+MEKAL model of the XRB determined from the SF region 
(figure \ref{fig:SrcFree} and table \ref{table:SrcFree})
was convolved with the rmf and the arf for the Lobe region, 
created in the manner similar to those of the SF region,
and plotted with the solid (PL) and dotted (MEKAL) lines.
The XRB model, convolved with the rmf and arf, 
yields a $0.7$ -- $7$ keV FI and BI count rate of 
$(1.63_{-0.10}^{+0.09}) \times 10^{-2}$ cts s$^{-1}$ and 
$(2.09_{-0.13}^{+0.11}) \times 10^{-2}$ cts s$^{-1}$, respectively.  
As a result, we have securely detected excess signals 
above the XRB level with an FI and BI count rate of  
$(0.83 \pm 0.05 _{-0.10}^{+0.11}) \times 10^{-2}$ cts s$^{-1}$
and 
$(1.18 \pm 0.09 _{-0.13}^{+0.15}) \times 10^{-2}$ cts s$^{-1}$
respectively.
Here, the first error represents the statistical error,
while the second error takes into account 
the error from the XRB model and the systematic error from the NXB. 
The residual $\chi$-spectrum in panel (a) of figure \ref{fig:image} 
clearly visualizes the significance of the excess 
($\chi^2/{\rm dof} = 594.9/97 $ in $0.7$ -- $7$ keV). 

We reproduced the excess emission from the east lobe 
by an additional PL component, subjected to the Galactic absorption 
($N_{\rm H} = 4.9 \times 10^{20}$ cm$^{-1}$). 
In order to create the arf for the lobe, 
we assumed a uniform emission filling a circle 
with a radius of $7$ arcmin ($293.8$ kpc at the source rest frame),
referring to the $608$ MHz radio image shown in figure \ref{fig:image}. 
As shown in panel (b) of figure \ref{fig:lobe},
a reasonable fit ($\chi^2/{\rm dof} = 117.6/95$)
was derived by the additional PL component 
with the parameters summarized in table \ref{table:lobe} (Case 1). 
The photon index, 
$\Gamma = 1.92_{-0.17}^{+0.13}$$_{-0.06}^{+0.04}$,
was found to be fairly consistent with that of 
the low-frequency synchrotron radio spectrum in the range of 326 -- 608.5 MHz,
$\Gamma_{\rm R} = 1.95$ \citep{GRG}.  
Here the first error is due to the photon statistics of the spectrum, 
while the second one is propagated from the systematic errors 
of both NXB and XRB. 
When the photon index was fixed at the radio index (Case 2),
we measured the 1 keV flux density of the additional PL component  
as $S_{\rm 1 keV} = 51.5\pm3.9_{-5.4}^{+6.2}$ nJy. 

\begin{table}[t]
\caption{Best-fit spectral parameters for the excess X-ray emission from the east lobe.}
\label{table:lobe}
\begin{center}
\begin{tabular}{lll}
\hline\hline 
Parameters              & Case 1                 & Case 2 \\
\hline 
$N_{\rm H}$ (cm$^{-2}$) & \multicolumn{2}{c}{$4.9 \times 10^{20}$ \footnotemark[$*$]}      \\
$\Gamma$ \footnotemark[$\ddagger$]               & $1.92_{-0.17}^{+0.13}$$_{-0.06}^{+0.04}$ & $1.95$ \footnotemark[$\dagger$]      \\
$S_{\rm 1 keV}$ (nJy) \footnotemark[$\ddagger$]   & $50.7_{-5.6}^{+4.7}$$_{-4.7}^{+5.3}$   & $51.5\pm3.9_{-5.4}^{+6.2}$\\ 
$\chi^2/{\rm dof}$      & $117.6/95$             & $117.9/96$\\ 
\hline 
\multicolumn{3}{@{}l@{}}{\hbox to 0pt{\parbox{75mm}{\footnotesize
\par\noindent
\footnotemark[$*$] Fixed at the Galactic value \citep{NH}.
\par\noindent
\footnotemark[$\dagger$] Fixed at the radio index \citep{GRG}. 
\par\noindent
\footnotemark[$\ddagger$] The first and second errors represent 
                          the statistical and systematic ones, respectively.
}\hss}}
\end{tabular}
\end{center}
\end{table}

\begin{figure}[t]
\begin{center}
\includegraphics[angle=-90,width=7.8cm]{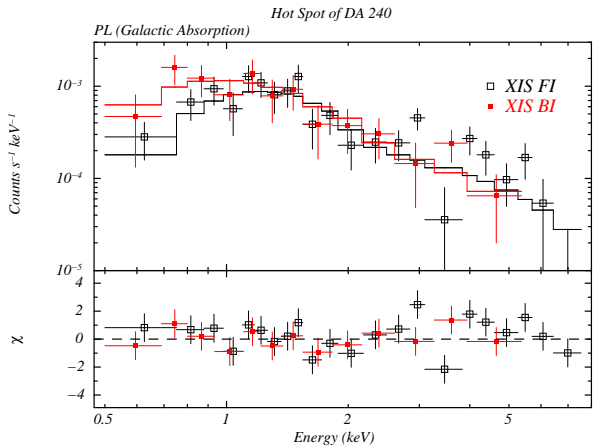}
\end{center}
\caption{Suzaku XIS spectrum of the east hot spot of DA 240,
which is successfully fitted with a simple PL model 
subjected to the Galactic absorption toward DA 240.}
\label{fig:hotspot}
\end{figure}

\subsection{X-ray spectra of the hot spot and nucleus} 
The X-ray signals were accumulated from the circle 
with a radius of $1'$ centered on the Chandra position of the hot spot.
The background was estimated 
from the $2'$ circle within the lobe, 
free from the X-ray point sources (shown in table \ref{table:CXO_Src}). 
The background-subtracted XIS spectrum of the east hot spot of DA 240 
is shown in figure \ref{fig:hotspot},
and the signal statistics are presented in table \ref{table:stats}.
The arf for a point source at the position of the hot spot is 
calculated with xissimarfgen.
We successfully described the observed spectrum 
with a simple PL model, modified with the Galactic absorption,
and tabulate the best-fit parameters in table \ref{table:hotspot}.
Both the photon index ($\Gamma = 1.99_{-0.27}^{+0.28}$) and 
1 keV flux density ($S_{\rm 1 keV} = 9.9\pm1.8$ nJy) are found to be consistent 
within the statistical uncertainties, with the XMM-Newton result 
($\Gamma = 2.2 \pm 0.3$ and $S_{\rm 1 keV} = 7\pm1$ nJy; \cite{DA240_HS}).

In the similar manner, the XIS signals from the nucleus of DA 240 
was integrated within the circle with a $1'$ radius centered on it,
while the background was derived from a neighboring point-source-free region 
with a $2'$ radius.
As we briefly stated in \S \ref{sec:image},
the anomalous columns of XIS 0 intersect the source region.
In addition, the nucleus is found to be unfortunately located 
at the boundary of the calibration source region for XIS 3.
These are possible to make a significant impact on a faint source.
Therefore, we decided to utilize only the XIS 1 (i.e., BI) data
for the nucleus.
As is shown in table \ref{table:stats},
we found that the nuclear XIS BI signals are statistically insignificant 
with a $3\sigma$ upper limit on the 0.7 -- 7 keV BI count rate of 
$5.6 \times 10^{-4}$ cts s$^{-1}$.
With the point source arf at the position of the nucleus, 
this is converted to the upper limit on 
the absorption-inclusive X-ray flux and luminosity at the source frame 
of $2.7 \times 10^{-14}$ ergs s cm$^{-2}$ 
and $8.0 \times 10^{40}$ ergs s$^{-1}$ in the $0.7$ -- $7$ keV range,
when we adopt the PL spectrum absorbed by the Galactic column density 
with a photon index of $\Gamma = 1.91$,
which is reported from the XMM-Newton observation 
of the DA 240 nucleus conducted on 2006 October 18 \citep{DA240_HS}.  
The 2 -- 10 keV X-ray intrinsic luminosity 
of $ 5.5 \times 10^{40}$ ergs s$^{-1}$ determined from the XMM-Newton spectrum, 
which corresponds to the 0.7 -- 7 keV absorption-inclusive luminosity 
of $7.1 \times 10^{40}$ ergs s$^{-1}$, 
is found to be consistent with the Suzaku upper limit. 
Thus, we detected no evidence of a significant intensity variation 
between the two observations, separated by $\sim 3.4$ years from each other. 

\begin{table}[t]
\caption{Best-fit spectral parameters for the hot spot.}
\label{table:hotspot}
\begin{center}
\begin{tabular}{ll}
\hline\hline 
Parameters                             & Values \\
\hline 
$N_{\rm H}$ (cm$^{-2}$)                 & $4.9 \times 10^{20}$ \footnotemark[$*$]      \\
$\Gamma$                               & $1.99_{-0.27}^{+0.28}$\\
$S_{\rm 1 keV}$ (nJy)                   & $9.9\pm1.8$ \\
$\chi^2/{\rm dof}$                     & $34.3/33$ \\ 
\hline 
\multicolumn{2}{@{}l@{}}{\hbox to 0pt{\parbox{80mm}{\footnotesize
\par\noindent
\footnotemark[$*$] Fixed at the Galactic value \citep{NH}.
}\hss}}
\end{tabular}
\end{center}
\end{table}

\section{Discussion} 
\label{sec:discussion}
\subsection{Energetics in the east lobe of DA 240} 
\label{sec:energetics}
With the careful subtraction of the XRB, NXB and 
contaminating X-ray point sources from the Suzaku XIS data,
we have revealed the significant X-ray emission 
from the east lobe of the giant radio galaxy DA 240.
We also detected the hot spot within the east lobe,
of which the XIS spectrum is reproduced by a PL model 
with a photon index of $\Gamma = 1.99_{-0.27}^{+0.28}$
and a 1 keV flux density of $S_{\rm 1 keV} = 9.9\pm1.8$ nJy),
although the nucleus was found to be too faint
with an upper limit on the 0.7 -- 7 keV observed luminosity of 
$8.0 \times 10^{40}$ ergs s$^{-1}$.

\begin{figure}[t]
\centerline{\FigureFile(80mm,80mm){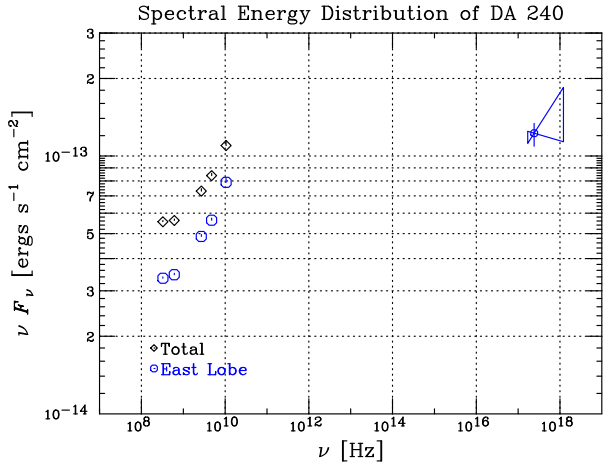}}
\caption{Spectral energy distribution of the east lobe of DA 240.
The synchrotron radio data \citep{GRG}
of the whole radio source and east lobe 
were plotted by the diamonds and circles, respectively. 
The best-fit PL component to the excess X-ray emission from the lobe 
was shown with the bow tie. }
\label{fig:sed}
\end{figure}

The XIS spectrum of the lobe was successfully described by a PL model 
with a photon index of $\Gamma = 1.92_{-0.17}^{+0.13}$$_{-0.06}^{+0.04}$.
We plotted the radio and X-ray spectral energy distribution 
of the east lobe in figure \ref{fig:sed}.
Here, we note that the contribution from the hot spot
was not subtracted from the radio data.
The high-frequency spectrum in the $4.8$ -- $10.6$ GHz range 
appears to be hard with a photon index of 
$\Gamma_{\rm R} = 1.58 \pm 0.01$ \citep{GRG},
since the hot spot is dominant in the range.
In contrast, the lobe exhibits a steep spectral index,
$\Gamma_{\rm R} = 1.95 \pm 0.01$ in the $326$ -- $608.5$ MHz,
where the contamination from the hot spot becomes unimportant \citep{GRG}.  
The agreement between the radio synchrotron and X-ray spectral slopes
of the lobe, in combination with the fact that 
the radio and X-ray spectra do not smoothly connect to each other
as we see in figure \ref{fig:sed}, 
strongly supports the interpretation that 
the synchrotron electrons in the east lobe radiate 
the excess X-ray emission through the IC scattering.
Such IC X-ray emission is widely observed from 
other radio sources (e.g., \cite{lobes_Croston}),
including giant radio galaxies (e.g., \cite{3C326,3C35}).

We evaluate the energy density 
of the possible seed photon candidates for the IC process in the lobe.
These include the CMB photons \citep{CMB_IC},  
infra-red (IR) radiation from the active nucleus \citep{IC_nuclearIR},
and the synchrotron photons produced within the lobe themselves.
Assuming an isotropic radiation, the IR flux density of the DA 240 nucleus, 
53 mJy at 60 $\mu$m \citep{IR_nucleus},
yields an IR photon energy density 
of $\sim 2 \times 10^{-16} (r/100{\rm~kpc})^{-2}$ ergs cm$^{-3}$
where $r$ is the distance from the nucleus.  
In the case of the FR II radio galaxies, including DA 240, 
IR photons from the base of the jet could serve as a seed photon source
to some extent. 
Since a large part of such IR emission is thought to be unobservable 
from our line of sight,
due to a beaming effect from the relativistic bulk motion of the jet,
it is not easy to evaluate precisely its energy density. 
Assuming a typical Doppler beaming factor of $\delta \sim 10$ 
(e.g., \cite{beaming_factor}),
the IC scattering off the jet IR photons is thought to become effective 
only within the narrow cone toward the jet direction
with an opening angle of $\sin \theta \sim 1/10$.
If this is the case,  the IC X-ray emission should be enhanced along the jet,
connecting the nucleus and hot spot. 
However, we do not find such spatial structure in the XIS image
in figure \ref{fig:image}. 
Therefore, we regard that such IC process is unimportant 
for the X-ray emission observed from the lobe of DA 240.
From the radio spectrum plotted in figure \ref{fig:sed},
the synchrotron photons are estimated to be negligible,
with an energy density of $\ll 10^{-18}$ ergs cm$^{-3}$
if it is spatially averaged over the lobe. 
Thus, we concluded that the CMB, 
with an energy density of $u_{\rm CMB} = 4.7 \times 10^{-13}$ ergs cm$^{-3}$
at the redshift of DA 240 ($z = 0.035661$),
highly dominates the other seed photon sources. 
The recent $\gamma$-ray result with Fermi from Centaurus A \citep{CenA_Fermi}
confirmed an additional contribution
from the extragalactic background light to the seed photon source.
However, such IC emission is only detectable 
at the highest energy end of the IC spectrum 
in the GeV $\gamma$-ray range with only a negligible X-ray flux.  

Based on the simple analytical formula 
for the IC/CMB process presented in \citet{CMB_IC},  
we derive the energy densities of the electrons and magnetic field,
$u_{\rm e}$ and $u_{\rm m}$ respectively, in the east lobe of DA 240.
The low-frequency radio synchrotron flux density and photon index
was adopted from \citet{GRG} as $S_{\rm R} = 10.30 \pm 0.12$ Jy at 326 MHz 
and $\Gamma_{\rm R} = 1.95 \pm 0.01$ in 326 -- 608.5 MHz.
Correspondingly, the electron number density spectrum was assumed 
to be a simple PL form as 
$\propto \gamma_{\rm e}^{-p}$ with $p = 2 \Gamma_{\rm R} -1 = 2.9$, 
where $\gamma_{\rm e}$ is the electron Lorentz factor. 
The IC/CMB X-ray flux density was measured from the PL fitting 
with the photon index fixed at this radio index, 
as $S_{\rm 1 keV} = 51.5\pm3.9_{-5.4}^{+6.2}$ nJy 
(Case 2 in table \ref{table:lobe}).
From the 608.5 MHz radio image shown in figure \ref{fig:image},
we approximate the shape of the lobe 
at a simple sphere with a radius of $ 293.8 \pm 4.2 $ kpc
(corresponding to $\timeform{7'} \pm \timeform{0.1'}$),
which gives a volume of $V = (3.13 \pm 0.13)\times 10^{72}$ cm$^3$.
The filling factor of the electron and magnetic field in the lobe 
were supposed to be unity.

\begin{figure}[t]
\centerline{\FigureFile(80mm,80mm){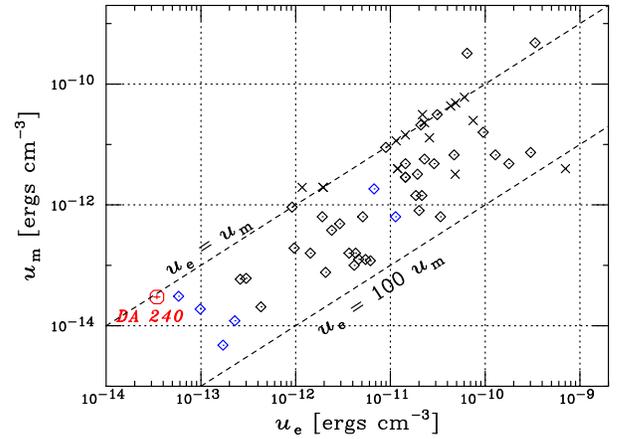}}
\caption{Summary of $u_{\rm e}$ and $u_{\rm m}$ in lobes of radio galaxies
(\cite{lobes_Croston,3C35,3C457_XMM}, and reference therein).
The east lobe of DA 240 is indicated by the red circle, 
with the statistical error.    
The lobes, from which the significant IC X-ray detection was reported, 
are plotted with diamonds,
while those with only the upper limit on the IC flux are shown with crosses.  
The giant radio galaxies are shown with the blue symbols. 
The two dashed lines represents the equipartition and a particle dominance 
of $u_{\rm e} / u_{\rm m} = 1 $, and $100$, respectively.}
\label{fig:ue-um}
\end{figure}

The energetics in the east lobe of DA 240 are summarized 
in table \ref{table:energetics}, 
together with the input quantities 
discussed above.
The energy densities of electrons and magnetic field were 
evaluated as 
$u_{\rm e} = (3.4_{-0.2}^{+0.3}$$_{-0.4}^{+0.5})\times 10^{-14}$
ergs cm$^3$ 
and 
$u_{\rm m} = (3.0\pm0.2\pm0.4) \times 10^{-14}$ ergs cm$^3$,
respectively;
the latter of which corresponds to the magnetic field 
of $B = 0.87\pm0.03_{-0.06}^{+0.05} $ $\mu$G.
Here, the first error represents the statistical error 
from $S_{\rm 1keV}$,
while the second one takes into account 
all the possible systematics 
from $S_{\rm 1keV}$, $S_{\rm R}$, $\alpha_{\rm R}$ and $V$.
We have confirmed an energy equipartition 
between the electrons and magnetic field
as $u_{\rm e}/u_{\rm m} = 1.1_{-0.1}^{+0.2}$$_{-0.2}^{+0.4}$. 

\subsection{Implication on the evolution of the lobe energetics} 
\label{sec:evolution}
In figure \ref{fig:ue-um} 
(\cite{lobes_Croston,3C35,3C457_XMM} and reference therein),
we compiled the relation between $u_{\rm e}$ and $u_{\rm m}$ 
in the lobes of radio galaxies, disentangled with the IC/CMB technique. 
Giant radio galaxies (shown with the blue symbols) tend 
to be distributed around the bottom-left corner in the diagram, 
except for the lobes of 3C 457 \citep{3C457_XMM}
of which the data point is located around the center of the distribution.
Especially, $u_{\rm e}$ in the east lobe of DA 240 is measured to be 
the lowest among all the X-ray detected lobes.  
In addition, $u_{\rm m}$ in the giant radio galaxies 
is typically indicated to be lower than than the CMB energy density. 
Thus, this result strengthened the idea that 
the dominance of the IC/CMB radiative losses over the synchrotron one 
is a common properties 
in the lobes of giant radio galaxies \citep{GRG_ICdominance}. 

As we frequently pointed out
in the previous paper (e.g., \cite{3C98,3C326,3C35}), 
the lobes typically exhibit an electron dominance 
of $u_{\rm e}/u_{\rm m} \sim 10$,
over a wide range of $u_{\rm e}$ and $u_{\rm m}$.
In contrast, 
a relatively large fraction of the giant radio galaxies 
detected with the IC/CMB X-rays 
(DA 240 and 3C 35 out of the 5 giants in total) 
is found to reside nearly in the equipartition condition.
We speculate that as the radio source develop 
from a $\lesssim 100$ kpc scale to a Mpc one, 
the energetics in their lobes evolve 
from the electron dominance to the equipartition, 
in addition to a significant decrease in both $u_{\rm e}$ and $u_{\rm m}$. 

\begin{figure}[t]
\centerline{\FigureFile(80mm,80mm){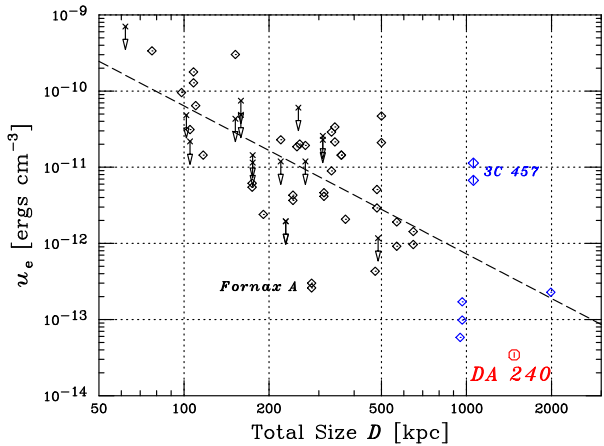}}
\caption{Electron energy density $u_{\rm e}$ in the lobes,  
plotted against the total physical size $D$ of the radio galaxies.
The east lobe of DA 240 is indicated with the red circle,
while the other giant radio galaxies are plotted with the blue diamonds.
The dashed line shows 
The best-fit $u_{\rm e}$--$D$ relation except for the giant radio galaxies.}
\label{fig:ue-D}
\end{figure}

The evolution of the lobe energetics is more explicitly 
examined in figure \ref{fig:ue-D},
which plots the relation between $u_{\rm e}$ 
and the total size $D$ of the radio galaxies.
It is pointed out that 
a radio galaxy develop along the relation of $u_{\rm e} \propto D^{-2}$
while its lobes are actively energized by its jets \citep{3C326,3C35}.
After the jet power supplied to the lobes become declined, 
the value of $u_{\rm e}$ is thought to be significantly reduced 
by adiabatic expansion and radiative losses due to 
synchrotron and IC/CMB emission. 
This idea is supported 
by a correlation of $u_{\rm e} \propto D^{-1.9 \pm 0.4}$
for the radio galaxies with $ D < 900 $ kpc (the black diamonds),
which is shown with the dashed line in figure \ref{fig:ue-D}. 
On the other hand, 
it is notable that 
for the radio galaxy with a dormant nucleus, Fornax A \citep{ForA_nucleus},
the data points are 
located below the correlation line by more than an order of magnitude
in figure \ref{fig:ue-D} \citep{ForA,ForA_Suzaku}.

\begin{table*}[t]
\caption{Summary of the energetics in the DA 240 east lobe.}
\label{table:energetics}
\begin{center}
\begin{tabular}{lll}
\hline\hline 
Parameters           & Values & Comment \\
\hline       
$S_{\rm 1keV}$ (nJy) & $51.5\pm3.9_{-5.4}^{+6.2}$  & Case 2 \\ 
$S_{\rm R}$    (Jy)  & $10.24 \pm 0.1 $            & 326 MHz  \\
$\Gamma_{\rm R}$     & $1.95 \pm 0.01 $            & 326 -- 608.5 MHz  \\
$V$ ($10^{72}$ cm$^3$) & $ 3.13 \pm 0.13 $         & \\
\hline       
$u_{\rm e}$ ($10^{-14}$ ergs cm$^{-3}$) \footnotemark[$*$]        
               & $3.4_{-0.2}^{+0.3}$$_{-0.4}^{+0.5}$ & $\gamma_{\rm e} = 10^3$ -- $10^5$\\
$u_{\rm m}$ ($10^{-14}$ ergs cm$^{-3}$) \footnotemark[$*$]        
               & $3.0\pm0.2\pm0.4$   &  \\
$B$ ($\mu$G) \footnotemark[$*$]  
               & $0.87\pm0.03_{-0.06}^{+0.05}$   &  \\
$u_{\rm e} / u_{\rm m}$ \footnotemark[$*$]
               & $1.1_{-0.1}^{+0.2}$$_{-0.2}^{+0.4}$   &  \\
\hline       
\multicolumn{2}{@{}l@{}}{\hbox to 0pt{\parbox{100mm}{\footnotesize
\par\noindent
\footnotemark[$*$] 
The statistics from $S_{\rm 1keV}$ is propagated to the first error,
while all the possible systematics are considered in the second one.  
}\hss}}
\end{tabular}
\end{center}
\end{table*}

Interestingly, the giant radio galaxies also tend to exhibit 
a lower value of $u_{\rm e}$ compared with the $u_{\rm e}$--$D$ relation 
for those with $ D < 900$ kpc. 
This implies a significant decrease 
in the jet activity of radio galaxies,        
in the range of $D = 100$ kpc--$1$ Mpc.
Even though the bright radio hot spots in DA 240 
suggesting an ongoing energy input to its lobes,  
the relatively steep synchrotron radio and IC X-ray spectra 
with the photon index of $\Gamma \sim 1.95$ indicate that 
the east lobe of this radio galaxy reside in a strong cooling regime. 
This means that the current energy supply to the lobe is ineffective,
in comparison with the cooling loss.  
In fact, the rather inactive nucleus of DA 240,
revealed in the XMM-Newton observation with an
observed luminosity of $7.1 \times 10^{40}$ ergs s$^{-1}$
and re-confirmed in this Suzaku observation 
with the upper-limit X-ray luminosity of 
$8.0 \times 10^{40}$ ergs s$^{-1}$ in the 0.7 -- 7 keV range, 
is though to be compatible with the currently low power jet scenario. 

In figures \ref{fig:ue-um} and \ref{fig:ue-D},
we found an outlier among the giant radio galaxies, 3C 457, 
with exceptionally high energy densities 
of $u_{\rm e} \sim 10^{-11}$ ergs s$^{-1}$ and 
$u_{\rm m} \sim 10^{-12}$ ergs s$^{-1}$ \citep{3C457_XMM},
which are comparable to those of the ``normal''
radio galaxies with $D\sim 200$ kpc. 
It is reported that 3C 457 hosts a rather active nucleus
with a 0.7 -- 7 keV intrinsic luminosity of 
$\sim 4 \times 10^{44}$ ergs s$^{-1}$ (estimated from \cite{3C457_XMM}),
which is more than 3 orders of magnitude higher than that of DA 240. 
Since its radio images revealed bright hot spots 
at the edge of the individual lobes \citep{3C457_XMM}, 
it is concluded that 
the lobes of 3C 457 is currently expanding very fast.
These are naturally regarded to be equivalent to 
the ongoing energy input to the lobes from the jets through the hot spots,
in this radio galaxy.
These active features of 3C 457 are thought to be related to 
the fact that the source is relatively young 
with a spectral age of $\sim 30$ Myr \citep{3C457_XMM}
as a giant radio galaxy (typically $\sim 100$ Myr; \cite{giants_age}).

Finally, based on the consideration above, 
we have proposed that the jet of typical radio galaxies is 
possible to reduce its activity, 
as they evolve in size typically from $D \sim 100$ kpc to $ D \sim 1$ Mpc,
although the number of giant radio galaxies,
studied with the IC/CMB X-ray emission, is still small.
In order to make a definite conclusion, 
X-ray observations of giant radio galaxies 
are strongly encouraged with Suzaku and
future high-sensitivity X-ray missions, 
including the next generation Japanese X-ray observatory 
ASTRO-H \citep{ASTRO-H}.

We are grateful to all the members of the Suzaku team,
for the successful operation and calibration.
We thank the referee for her/his kindness to improve the paper. 
This research has made use of the archival Chandra data and
its related software provided by the Chandra X-ray Center (CXC).
The support is acknowledged from the Ministry of Education, Culture, Sports, 
Science and Technology (MEXT) of Japan 
through the Grant-in-Aid for the Global COE Program,
"The Next Generation of Physics, Spun from Universality and Emergence".
This investigation is partially supported by the
MEXT Grant-in-Aid for Young Scientists (B) 22740120 (N. I.) 
and for Scientific Research (B) 22340039 (M. S. T.).



\begin{thebibliography}{}
\bibitem[Abdo et al.(2010)]{CenA_Fermi}
  Abdo, A.A., et al., 2010, Science, 328, 725
\bibitem[Brunetti et al.(1997)]{IC_nuclearIR}
  Brunetti, G., Setti, G., \& Comastri, A.,
  1997, \aap, 325, 898
\bibitem[Croston et al.(2005)]{lobes_Croston}
  Croston, J. H., Hardcastle, M. J., Harris, D. E., 
  Belsole, E., Birkinshaw, M., \& Worrall, D. M.,
  2005, \apj, 626, 733
\bibitem[Ebneter \& Balick(1985)]{DA240_host}	
  Ebneter, K., \&  Balick, B.,
  1985, \aj, 90, 183
\bibitem[Evans et al.(2008)]{DA240_HS}
  Evans, D. A., Hardcastle, M. J., Lee, J. C., Kraft, R. P.,
  Worrall, D. M., Birkinshaw, M., \& Croston, J. H.,
  2008, \apj, 688, 844
\bibitem[Feigelson et al.(1995)]{ForA_ROSAT}
  Feigelson, E. D., Laurent-Muehleisen, S. A.,
  Kollgaard, R. I., \& Fomalont, E. B.,
  1995, \apj, 449, L149 
\bibitem[Golombek et al.(1998)]{IR_nucleus}
  Golombek, D., Miley, G. K., \& Neugebauer, G.,
  1988,\aj, 95, 26
\bibitem[Harris \& Grindlay(1979)]{CMB_IC}
  Harris, D. E., \&  Grindlay, J. E.,
  1979, \mnras, 188, 25
\bibitem[Inoue \& Takahara(1996)]{beaming_factor}
  Inoue, S. \& Takahara, F., 1996, \apj, 463 555,
\bibitem[Ishisaki et al.(2007)]{xissimarf}
  Ishisaki, Y.,  et al. 2007, \pasj, 59, 113
\bibitem[Ishwara-Chandra \& Saikia(1999)]{GRG_ICdominance}
  Ishwara-Chandra, C. H., \& Saikia, D. J.,
  1999, \mnras, 309, 100
\bibitem[Isobe et al.(2002)]{3C452}
  Isobe, N., et al., 2002, \apj, 580, L111
\bibitem[Isobe et al.(2009)]{3C326}
  Isobe, N., et al., 2009, \apj, 706, 454
\bibitem[Isobe et al.(2005)]{3C98}
  Isobe, N., Makishima, K., Tashiro, M., \& Hong, S., 2005, \apj, 632, 781
\bibitem[Isobe et al.(2006)]{ForA}
  Isobe, N., Makishima, K., Tashiro, M., Itoh, K.,
  Iyomoto, N., Takahashi, I., \& Kaneda, H., 2006, \apj, 645, 256
\bibitem[Isobe et al.(2011)]{3C35}
  Isobe, N., Seta, H., Gandhi, P., \& Tashiro, M.S., 
  2011, \apj, 727, 82
\bibitem[Iyomoto et al.(1998)]{ForA_nucleus}
  Iyomoto, N., Makishima, K., Tashiro, M., Inoue, S.,
  Kaneda, H., Matsumoto, Y., \& Mizuno, T.
  1998, \apj, 503, L31
\bibitem[Kalberla et al.(2005)]{NH}
  Kalberla, P. M. W., Burton, W. B., Hartmann, Dap, 
  Arnal, E. M., Bajaja, E., Morras, R., \& Po\"{o}ppel, W. G. L.,
  2005,\aap, 440, 775
\bibitem[Kaneda et al.(1995)]{ForA_ASCA}
  Kaneda, H., et al., 1995, \apj, 453, L13  
\bibitem[Konar et al.(2010)]{3C457_XMM}
  Konar, C., Hardcastle, M. J., Croston, J. H., \& Saikia, D. J.
  2010, \mnras, 400, 480
\bibitem[Koyama et al.(2007)]{XIS}
  Koyama K., et al., 2007, \pasj, 59, S23
\bibitem[Kushino et al.(2002)]{XRB_ASCA}
  Kushino, A., Ishisaki, Y.,  Morita, U., Yamasaki, N. Y.,
  Ishida, M., Ohashi, T., \& Ueda, Y., 
  2002, \pasj, 54, 327
\bibitem[Laing et al.(1983)]{RG_summary}
  Laing, R. A., Riley, J. M., \& Longair, M. S.
  1983, \mnras, 204, 151
\bibitem[Lumb et al.(2002)]{XRB_XMM}
  Lumb, D. H., Warwick, R. S., Page, M., \& De Luca, A., 
  2002, \aap, 389, 93  
\bibitem[Mack et al.(1997)]{GRG}
  Mack, K.-H., Klein, U., O'Dea, C.P., \& Willis, A.G.,
  1997, \aaps, 123, 423
\bibitem[Mewe et al.(1985)]{MEKAL}
  Mewe, R., Gronenschild, E. H. B. M., \& van den Oord, G. H. J.
  1985, \aaps, 62, 197
\bibitem[Mitsuda et al.(2007)]{Suzaku}
  Mitsuda, K., et al., 2007, \pasj,  59, S1 
\bibitem[Rines et al.(2000)]{redshift}
  Rines, K., Geller, M. J., Diaferio, A.,
  Mohr, J. J., \& Wegner, G. A.,
  2000, \aj, 120, 2338
\bibitem[Serlemitsos et al.(2007)]{XRT}
  Serlemitsos, P.J, et al., 2007, \pasj, 59, S9
\bibitem[Schoenmakers et al.(2000)]{giants_age}
  Schoenmakers, A. P., Mack, K.-H., de Bruyn, A. G.,
  R\"{o}ttgering, H. J. A., Klein, U.,\& van der Laan, H.,
  2000, \aaps, 146, 293
\bibitem[Strom et al.(1981)]{DA240_age}
  Strom, R.G., Baker, J.R., \& Willis, A.G.,
  1981, \aap, 100, 220
\bibitem[Takahashi et al.(2007)]{HXD}
  Takahashi, T., et al., 2007, \pasj, 59, S35
\bibitem[Takahashi et al.(2010)]{ASTRO-H}
  Takahashi, T., et al., 2010, SPIE, 7732, 27 
\bibitem[Tashiro et al.(1998)]{CenB}
  Tashiro, M., et al., 1998, \apj, 499, 713
\bibitem[Tashiro et al.(2009)]{ForA_Suzaku}
  Tashiro M., Isobe, N., Seta H., Yaji, Y., \& Matsuta K., 
  2009, \pasj, 61, S327
\bibitem[Tashiro et al.(2001)]{ForA_ASCA_2}
  Tashiro, M., Makishima, K., Iyomoto, N., Isobe, N., \& Kaneda, H.,
  2001, \apj, 546, L19
\bibitem[Tawa et al.(2008)]{xisnxbgen}
  Tawa, N., et al., 2008, \pasj, 60, S11
\bibitem[Uchiyama et al.(2008)]{Suzaku_pos_acc}
  Uchiyama et al., 2008, \pasj, 60, S35
\bibitem[Yaji et al.(2010)]{CygA_Yaji}
  Yaji, Y., et al., 2010, \apj, 714, 37
\end{thebibliography}
\end{document}